\documentclass[11pt]{article}
\usepackage{moriond,epsfig}

\def\be{\begin{equation}}
\def\ee{\end{equation}}
\def\bea{\begin{eqnarray}}
\def\eea{\end{eqnarray}}

\begin{document}
\vspace*{4cm}
\title{ELLIPTIC FLOW AND SEMI-HARD SCATTERING AT SPS}

\author{J.P. WURM and J. BIELCIKOVA$^1$\\
for CERES/NA45}

\address{Max-Planck-Institut f\"ur Kernphysik, Saupfercheckweg, 
69117 Heidelberg\\ and Physikalisches Institut der Universit\"at
Heidelberg, 69120 Heidelberg\\
$^1$ Present address: Nuclear Structure Lab., Yale University, 
New Haven, Ct. 06511, USA \\ }

\maketitle\abstracts{Results on elliptic flow and two-particle
correlations in the semi-hard regime are presented.}

\section{Introduction}
\indent

The first rounds of experiments at the Relativistic Heavy Ion Collider
(RHIC) at BNL have provided us with several clues\cite{qm04}: the
collective expansion following the violent collision of two Au nuclei
at $\sqrt s=$ 130 and 200 GeV energy per nucleon appears to resemble
the outward flow of a perfect fluid. This has been concluded from the
fact that the anisotropic, or elliptic, flow measured as the second
Fourier coefficient of the particle distribution with respect to the
event plane is very large and even reaches up to the predictions of
ideal, i.e. {\it non-viscous}, hydrodynamics. Moreover, elliptic flow,
commonly regarded as a property of particles from 'soft' processes,
characterized by a scale $\Lambda^2\ll$ 1 GeV$^2$, persists at large
$p_T$. This seems to be linked to another highlight from RHIC, the
suppression of hadron spectra at high $p_T$ (in comparison to the
cumulative yield of nucleon-nucleon collisions) together with the
disappearance of back-to-back correlations of such hadrons.  Both are
attributed to {\em jet quenching} in the Quark Gluon Plasma in which
partons after an initial hard encounter loose energy, mostly by gluon
radiation. The quenching of large $p_T$ hadrons is believed to be at
the origin of anisotropic flow in the semi-hard or hard sector as
observed by large azimuthal anisotopies.

We will report here on measurements\cite{prl} of inclusive particle
$p_T$ spectra, elliptic flow and azimuthal particle correlations up to
$p_T\approx$ 3 GeV/$c$ in Pb-Au collisions at the CERN Super-Proton
Synchrotron (SPS) at $\sqrt s$= 17.3 GeV.  We shall elaborate on those
issues in which SPS physics appears to be at variance with that at
RHIC.

\section{Inclusive spectra}
\indent

Pions are identified by the double Cherenkov ring imaging spectrometer
($\gamma_{thr}\approx 32$), their momenta measured by the ring radii.
Charged hadrons are tracked by a Si-drift telescope and a
\begin{figure}[t!]
\epsfig{figure=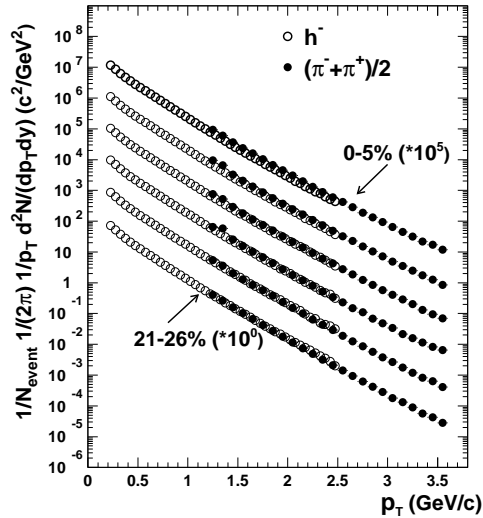,height=7.9cm}

\vspace{-4.7cm}

\begin{flushright}
 \begin{minipage}{0.5\linewidth} 
\caption{\label{inclusive} Invariant $p_T$ spectra of 
inclusive charged hadrons and pions. 158 GeV/$c$ Pb-Au, 
2.1$\leq y \leq$ 2.65.
Different centralities labeled by the percentage of the top inelastic 
cross section. The lowest, most peripheral spectrum (21-26$\%$, 
n$_{coll}$= 293) is at large $p_T$ steeper than the most central
spectrum at the top (0-5$\%$, 
$n_{coll}$= 774), indicating moderate Cronin enhancement above 2.5 GeV/$c$,
see text.}
\end{minipage}
\end{flushright}
\end{figure}
multi-wire chamber sandwiching the Cherenkov detectors,
with magnetic deflection inbetween. 

Invariant $p_T$ spectra (Fig.\ref{inclusive}) of negatively charged
hadrons, $h^-$, and of $\pi^\pm$ are found in close agreement despite
entirely different methods used. The concave shape above about
1.5 GeV/$c$ might indicate the power-law characteristics of
elementary parton interactions. Inverse slope parameters increase
monotonically from 180 to 270\,MeV/$c$ between $p_T$= 0.5 and
2.5\,GeV/$c$, respectively, in close agreement with $\pi^\circ$ 
data\cite{wa98} of WA98.

Reanalyzing the WA98 data using a better p-p reference,
d'Enterria\cite{denterria} finds only weak Cronin
enhancement\cite{cronin} and even some suppression of yields at
high-$p_T$ in most central 158 A\,GeV Pb\,Pb collisions, in contrast to
previous analyses\cite{wa98,wang98}. In lack of a suitable p-p reference
for $\pi^\pm$, we compare here the most central to the most peripheral
spectrum using the ratio
\begin{equation}
R_{CP}(p_T)= 
\frac{\langle d^2N/dp_Tdy\rangle/\langle n_{coll}\rangle(central)}
{\langle d^2N/dp_Tdy\rangle/\langle n_{coll}\rangle(peripheral)}.
\end{equation}
For the CERES data of Fig.\ref{inclusive}, $R_{CP}$ increases from
0.75 to 1.25 for $p_T$ between 1.3 and 3.5 GeV/c, respectively.  So
there is very moderate Cronin enhancement at large $p_T$ and no
indication of suppression. A possible alternative could be that both
effects are present, but nearly cancel.

\section{Elliptic flow}

\begin{figure}[t!]
\epsfig{figure=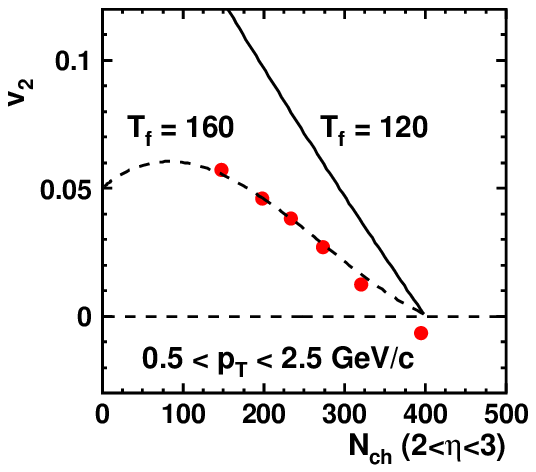,height=5.8cm}
\epsfig{figure=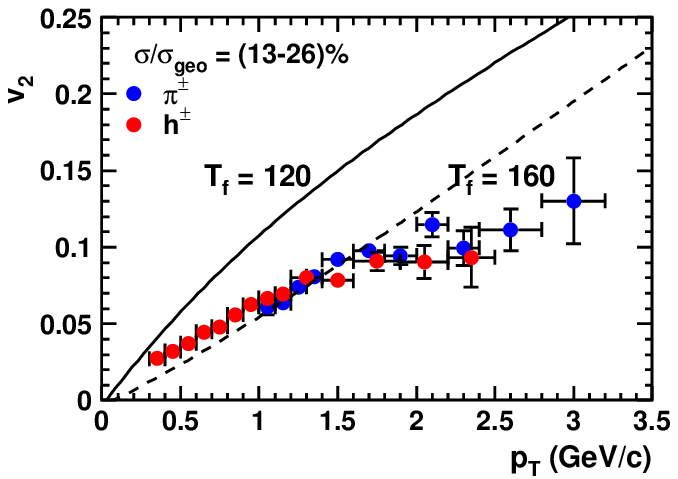,height=5.8cm}
\caption{\label{v2-compo} $v_2$, dependence on centrality (left, for $h^-$)
 and transverse momentum (right, for $h^-$ and $\pi^\pm$), compared to
 ideal hydrodynamical calculations of P. Huovinen using an equation of state
with a phase transition at 165 MeV and freeze-out at 120 and 160 MeV.}
\end{figure}

The results of the elliptic flow measurement are shown in
Fig.\ref{v2-compo}. The $v_2$ values have been corrected for the
measured resolution in determining the event plane using independent
particle samples and for effects of Bose-Einstein correlations of
particles close in phase space. Both the decrease with centrality and
the linear rise with $p_T$ might be taken to resemble perfect
hydrodynamical behaviour.  However, the calculations overshoot the
data by a large margin, and this deficiency apparently is not easily
cured by enforcing early freeze-out at larger $T_f$= 160 MeV since
proton spectra (not shown) are no longer described. It has been
suggested that inclusion of viscosity will considerably reduce
$v_2$\cite{bass,teaney}.

The rise of $v_2(p_T)$ gives way to saturation around
2 GeV/$c$, a feature quite similar to that observed at
RHIC\cite{snellings03}. This is the more surprising as $v_2$ at SPS
stays well below the non-viscous hydrodynamical limit. Yet, it is somewhat
puzzling that hydrodynamic calculations for $v_2$ at RHIC figure 
{\it below} those for the top SPS energy.

\section{Two-particle correlation in azimuth: collective and partonic}

We turn to the opening angle distribution in azimuth (in the plane
transverse to the beam), $dN/d\Delta\phi(\Delta\phi)$, between any two
high-$p_T$ pions in a given event. The anisotropy is measured by
the harmonic coefficient $p=
\langle\,cos(2\,\Delta\phi)\,\rangle$. {\it Direct} correlations cause
the anisotropy to exceed the pure-flow magnitude $\sqrt p= v_2$.
\begin{figure}[b!]
\epsfig{figure=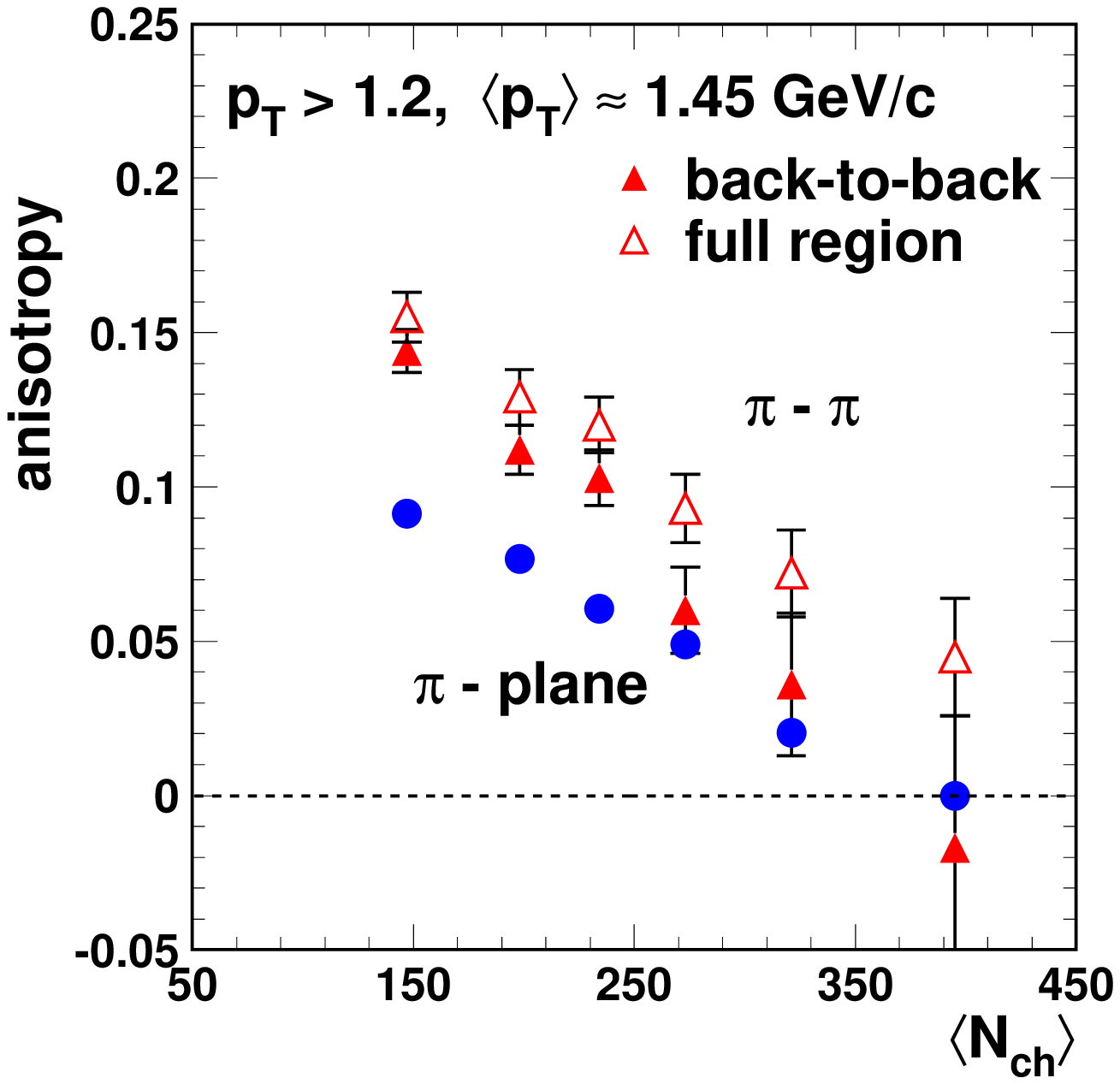,height=6.6cm}
\epsfig{figure=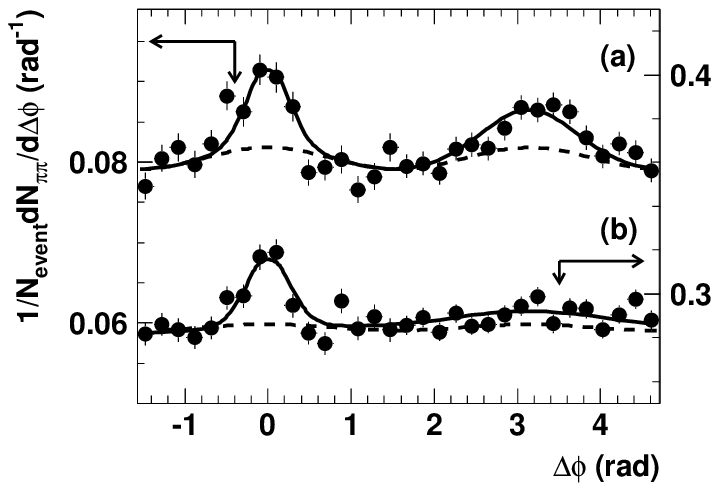,height=6.5cm}
\caption{\label{aniso-gauss} Left: Centrality dependence of 
anisotropy parameters $\sqrt p$ from $\pi\pi$ correlations (triangles)
and $v_2$ from standard flow measurement (circles). Full triangles
give anisotropy for a restricted range $\pi/2\leq \Delta\phi
\leq\,3\,\pi/2$. Right: $\Delta\phi$ distribution
for pairs of pions of $p_T\geq$\,1.2\,GeV/$c$, corrected for detection
efficiency. Centrality (a) 21-26$\%$, (b) 9-13$\%$.  The
$\Delta\phi$=\,0 region was protected against distorsions by invoking
a minimum track separation $\Delta\theta\geq$20\,mr.  The dotted line
denotes the independently measured flow contribution, the full line is
a fit with two Gaussians fixed at $\Delta\phi$= 0 and $\pi$. }
\end{figure}
As seen from Fig.\ref{aniso-gauss}, the standard $v_2$ accounts for only
about 60$\%$ of $\sqrt p$, the anisotropy in opening angle, the
remaining {\it non-flow} component is presumably of semi-hard
origin. To investigate dijet-like correlations, the measured
$\Delta\phi$ distribution (Fig.\ref{aniso-gauss} r.h.s.) is described
by two Gaussians centered at $\Delta\phi$= 0,\,$\pi$ on top of the
measured elliptic flow. 

The peaks correspond to average momentum transfers $Q^2= (2\,\langle
p_T \rangle)^2\approx$ 8.4 GeV$^2$, well in the semi-hard regime;
the invariant mass of the back-to-back 'object' is above 2 GeV/$c^2$
which makes resonance decays quite unlikely. 
\begin{figure}[t!]
\epsfig{figure=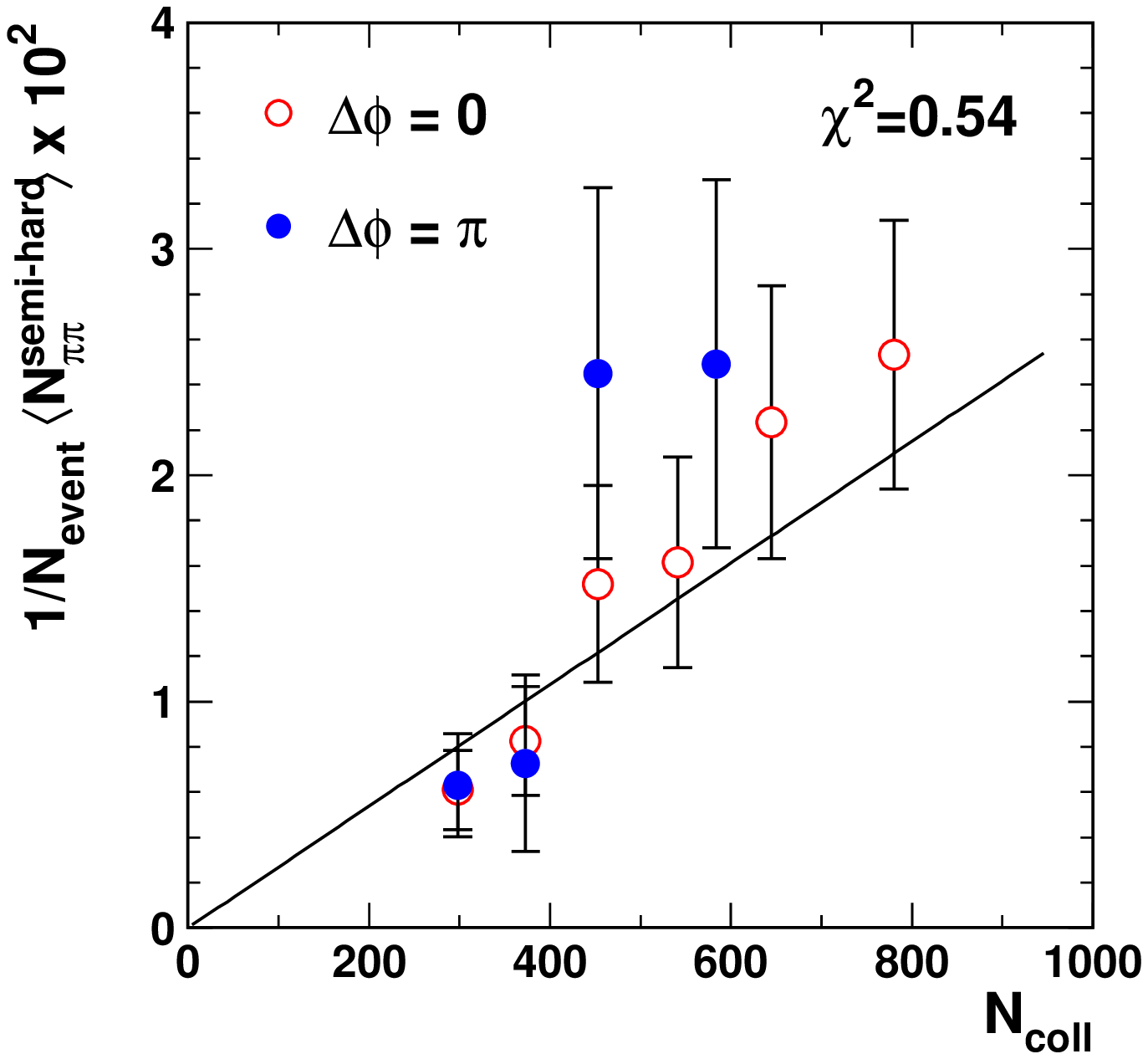,height=6.0cm}
\epsfig{figure=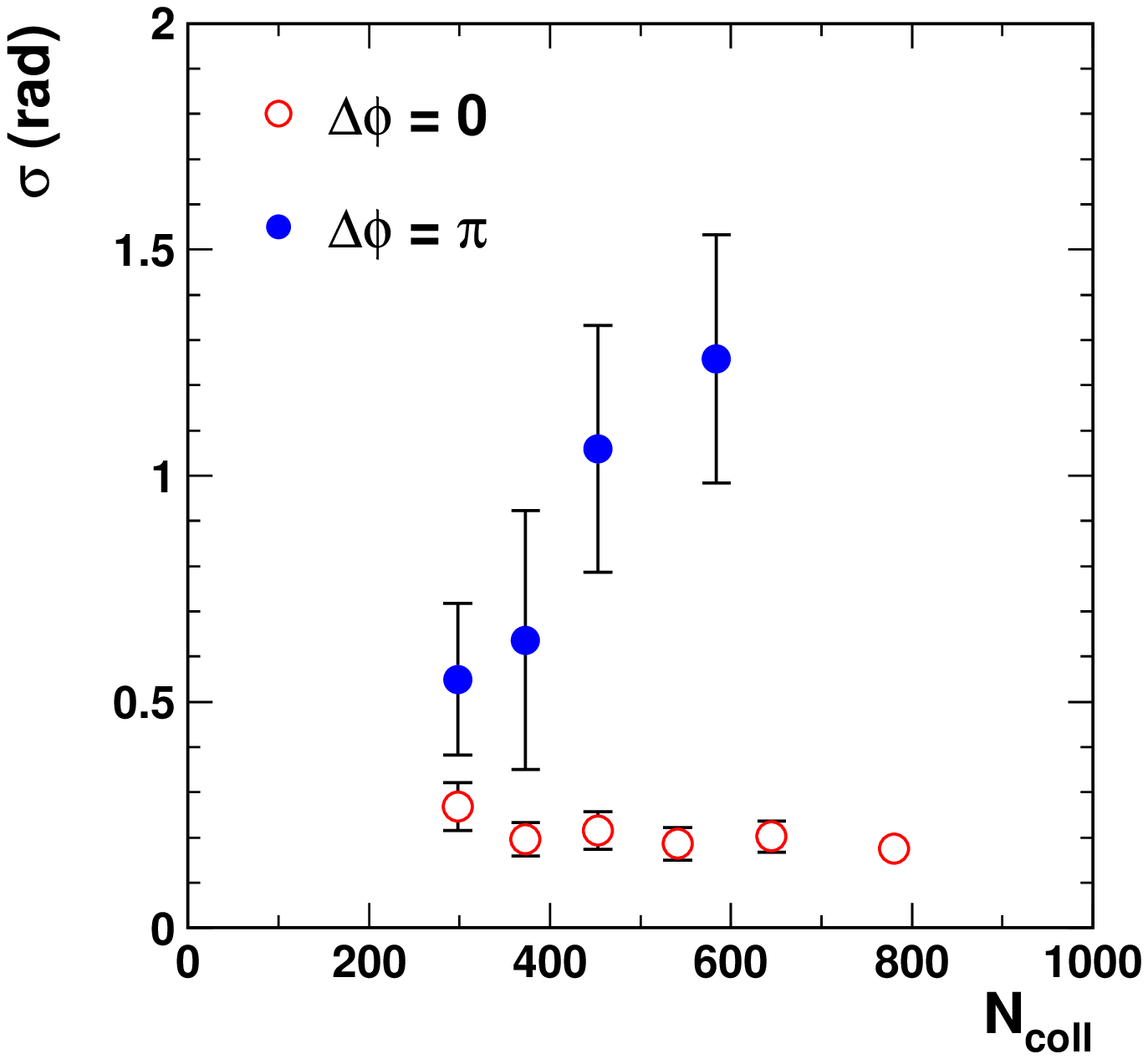,height=6.0cm}
\caption{\label{yield-sigma-binary} Centrality dependence of semi-hard
yields (left) and widths of Gausian peaks (right). The 'back-to-back'
component is lost in background for $N_{coll}>$\,600. The $\chi^2$
of straight-line fit doubles if plotted {\it vs} $N_{part}$ instead.}
\end{figure}
Identifying the close-angle peak with parton fragmentation, its width
$\sigma_0$= (0.23$\pm$0.03)\,rad corresponds to an average pion
momentum perpendicular to the jet thrust axis of $\langle
j_T^2)^{1/2}= \sqrt{2}\langle p_T\rangle\,\sigma_0$=
(472$\pm$62)\,MeV/$c$. This agrees with p-p results from
ISR\cite{angelis} and RHIC\cite{rak}. The yields of the 'near-angle' and 
'back-to-back' peaks\footnote{Symmetric parton-parton scattering within
CERES acceptance requires a partonic reference frame moving backward
with $p_L\approx$\,1.5 GeV/$c$ corresponding to $x_F$=\,--0.2.}
grow linearly with the number of binary collisions,
Fig.\ref{yield-sigma-binary}; yet, the back-to-back component
disappears in more central collisions because its width increases
rapidly (see r.h.s. of Fig.\ref{aniso-gauss}, b).

Of course, the excitement is with the disappearance of the
back-to-back component. We derive a $k_T$ broadening of
(2.8$\pm$0.6)\,GeV/$c$ for the most-central data point which is
significantly enlarged by the nuclear medium compared to
p-p\cite{angelis} and d-Au collisions\cite{rak}, and quite similar
to preliminary results from RHIC\cite{miller} for Au-Au. The fact
that we see no indication of suppression might be owed to the rather
low $p_T$ threshold so that quenched hadrons are still
accepted\cite{fwang}; these and other open questions are currently
investigated in more detail\cite{aga_new}.

\end{document}